\documentclass[12pt]{article}
\usepackage{bbm,latexsym}
\usepackage{amssymb,amsmath}
\usepackage{slashed}
\newcommand{\dd}{\mathrm{d}}
\newcommand{\ma}{\mathcal{A}_{\nu_\alpha \to \nu_\beta}}
\newcommand{\mt}{\mathcal{T}}
\newcounter{assumption}
\allowdisplaybreaks
\begin{document}
\title{
\normalsize \hfill UWThPh-2023-10 \\[10mm]
\LARGE Yet another QFT model of neutrino oscillations}

\author{
W.~Grimus\thanks{E-mail: walter.grimus@univie.ac.at}\;
\addtocounter{footnote}{1}
\\[5mm]
\small University of Vienna, Faculty of Physics \\
\small Boltzmanngasse 5, A--1090 Vienna, Austria
}

\date{April 5, 2023}

\maketitle

\begin{abstract}
We consider a quantum field theory 
(QFT) model of neutrino oscillations in vacuum that attempts 
to take into account that the neutrino source particle and the neutrino 
detection particle both interact with their respective environments by 
collisions. Our model is minimal in a twofold sense. Firstly 
we simply assume that  neutrino production and detection take place 
in between collisions in time intervals $\Delta t$ and $\tau$, respectively. 
Secondly, we only introduce the two wave packets that are absolutely necessary 
which are those of the neutrino source and the neutrino 
detection particle. Within this model we find that, for all practical 
purposes, there are no decoherence effects in the neutrino oscillation 
amplitude and oscillations occur in space not in time. Moreover, our model 
leads to the correct time correlation between neutrino production 
and detection and to a factorization of the event rate of the compound 
neutrino production--detection process into decay rate, oscillation 
probability and detection cross section.
\end{abstract}

\newpage

\section{Introduction}
QFT is a suitable framework for describing 
neutrino oscillations---for reviews see for instance~%
\cite{bilenky-rev,petcov-rev,zralek,giunti-rev,beuthe,G2003,akhmedov,athar}, 
for recent literature we refer the reader to~%
\cite{akhmedov2009,boyanovsky1,boyanovsky2,grimus2019,akhmedov2022,
simkovic,schwetz} and references therein.
In particular, QFT makes allowance for the fact 
that neutrinos are not directly observable. This is accomplished by ascribing 
a single Feyman diagram to the compound process of neutrino production 
and detection in which the massive neutrinos are inner lines.

In this paper we introduce a minimal QFT model that tries to take into 
account that neutrino production and detection do usually not take 
place in vacuum but in an environment. This is true for both reactor 
and accelerator neutrinos that we have in mind in this context. In a reactor 
the fission fragments that produce neutrinos by $\beta$-decay are in a thermal 
environment. We adopt the simple picture that this decay may occur 
without interruption during a time interval $\Delta t$. Thus $\Delta t$ 
is the typical time between two collisions. If a fission 
fragment has not decayed during this interval, it will undergo a collision 
with a surrounding atom and the process starts anew. Similarly, in the 
detection process we denote the time interval of uninterrupted measurement 
by $\tau$. Possibly, $\tau$ can also 
be interpreted as the time resolution of the 
detector. Note that we assume that the detector particle is at rest apart 
from thermal motion and is embedded in some kind of gas or liquid. 
Accelerator neutrinos are generated by the decay of charged pions in a 
decay tunnel which is the environment in this case. Here $\Delta t$ is the 
time between the pion creation and the moment when its trajectory intersects 
the end of the decay tunnel. The notation concerning the different times 
used in this paper is explained in table~\ref{table}.
\begin{table}
\renewcommand{\arraystretch}{1.3}
\begin{center}
\begin{tabular}{l|l}
Symbol & Meaning \\ \hline
$t_S$ & Time of creation of the source particle \\
$\Delta t$ & Time interval of source particle decay given by  
$(t_S, t_S + \Delta t)$ \\
$T$ & Time that occurs in $\ma$ after integrations \\
$t_D$ & Time of measurement \\
$\tau$ & Time interval of measurement given by $(t_D - \tau/2, t_D + \tau/2)$
\\
$\lambda$ & Time parameter that occurs in the integral over the neutrino energy
\\
$t_c$ & Time parameter used to describe the correlation between $t_S$ and $t_D$
\\
$\mt(t_c)$ & Temporal function that occurs in the decay rate 
\end{tabular}
\end{center}
\caption{List of time symbols used in the paper. $T$ is defined in 
equation~(\ref{T}) and $\ma$ denotes the oscillation amplitude referring 
to neutrino flavours $\alpha$ and $\beta$. The parameters $\lambda$ and $t_c$ 
are defined in equation~(\ref{lt}) and the function $\mt(t_c)$ in 
equation~(\ref{tc}).}
\label{table}
\end{table}

Our model is also minimal with respect to the number of wave packets. Only 
wave packets for the neutrino source particle and the detector particle 
are introduced. We assume plane waves for all particles in the final states, 
which is the usual practice in QFT computations of decay rates and cross 
sections. Obviously, there are no neutrino wave packets since neutrinos 
correspond to inner lines in the Feynman diagram of the compound process.

The subjects of the paper are 
\begin{itemize}
\item
the computation of the neutrino oscillation amplitude $\ma$ in our model,
\item
potential decoherence effects in $\ma$,
\item
and the factorization of the event rate into (decay rate of the source 
particle) times (oscillation probability) times (detection cross section).
\end{itemize}

The paper is organized as follows. 
In section~\ref{amplitude} the oscillation amplitude $\ma$ is derived, 
starting from a preparatory form discussed in appendix~\ref{integrations}. 
The integration over the neutrino energy in some suitable approximation 
is done in section~\ref{Q}. A general integral needed in this context is 
computed in appendix~\ref {integral}. Section~\ref{time} treats the 
correlation between the times $t_S$, when the source particle is created, 
and $t_D$, when the neutrino detection takes place; this correlation 
results from the 
integration over the neutrino energy. The connection between this 
time-correlation function and approximate energy conservation is established 
in section~\ref{function}. The above-mentioned factorization is deduced 
in section~\ref{accelerator} and the conclusions are presented in 
section~\ref{concl}.

\section{Oscillation amplitude}
\label{amplitude}
The space-time variable at the neutrino source process is denoted by $x_1$ 
and that at the detection process by $x_2$. We describe the neutrino source and 
detector particles by the wave functions
\begin{equation}\label{psiS}
\int \dd^3 p\, e^{-i( x_1 - x_S ) \cdot p}\, 
\psi_S(\vec p\,) \,e^{-\Gamma m_S(t_1 - t_S)/(2 E_S)}\, \Theta( t_1 - t_S )
\Theta( t_S + \Delta t - t_1)
\end{equation}
and
\begin{equation}\label{psiD}
\int \dd^3 k\, e^{-i( x_2 - x_D ) \cdot k}\, \psi_D(\vec k\,)
\,\Theta_\tau( t_2 - t_D ),
\end{equation}
respectively. For later convenience both wave functions are given as 
Fourier transforms. The vector $\vec x_S$ shifts the peak of 
the source wave function from $\vec x_1 = \vec 0$ to $\vec x_1 = \vec x_S$, 
with the analogue $\vec x_D$ for the detection wave function. 
The 4-momenta $p$ and $k$ are decomposed into time and space components as
\begin{equation}
p = \left( \begin{array}{c} E_S \\ \vec p \end{array} \right)
\quad \mbox{and} \quad
k = \left( \begin{array}{c} E_D \\ \vec k \end{array} \right),
\end{equation}
respectively. We denote the mass of the source particle by $m_S$ 
and that of the detector particle by $m_D$.

The Heaviside function $\Theta( t_1 - t_S )$ and the exponential---with 
the total decay rate $\Gamma$---describing 
the decay of the source particle are incorporated into 
the source wave function; this means that $t_S$ is the time when the source 
particle is created. For instance, for reactor neutrinos this is 
the time when the nucleus that undergoes $\beta$-decay is created by fission. 
As described in the introduction, we further assume that the decay process 
of the source particle is stopped at a time $t_S + \Delta t$ 
which is taken into account by $\Theta( t_S + \Delta t - t_1)$. 

The function $\Theta_\tau( t_2 - t_D )$, which we incorporated into the 
wave function of the detector particle, 
specifies the uncertainty $t_D \pm \tau/2$ in the detection time. 
The simplest form of $\Theta_\tau( t )$ is rectangular:
\begin{equation}\label{Thetatau}
\Theta_\tau( t ) = \left\{ 
\begin{array}{cl} 1 & \mbox{for}\; |t| < \tau/2, \\
0 & \mbox{otherwise.} 
\end{array} \right.
\end{equation}
Note that in the case of reactor neutrinos measured by the inverse 
$\beta$-decay the time $t_D$ is the time of the prompt signal.

We do not need to specify the dispersion relations
\begin{equation}
p^0 \equiv E_S(\vec p\,), \quad k^0 \equiv E_D(\vec k\,).
\end{equation}
They could be that of a free particle or something else. 
The derivations
\begin{equation}
\vec \beta_S = \vec\nabla_p E_S(\vec p\,), \quad 
\vec \beta_D = \vec\nabla_k E_D(\vec k\,)
\end{equation}
are the group velocities of source and detector particle, 
respectively.

To keep track of the assumptions we use in 
the computation of the oscillation amplitude $\ma$ we number them.
The first one is the following.
\begin{quote}
\refstepcounter{assumption}\label{fi}
{\bf Assumption~\arabic{assumption}:} The final states in both the neutrino 
source and detection process are treated as plane waves.
\end{quote}
This is the usual assumption for computing cross sections or decay rates 
in particle physics. This means that we do not consider possible 
effects of wave packets of the final states. Note, however, that 
such wave packets, and also $\psi_S$ and $\psi_D$, are not to be confused with 
wave functions of composite particles such as nuclei, nucleons, etc. 
Such particles could be described by some bound-state wave functions 
pertaining to their inner structure, however, the inner structure would 
be accounted for in the weak-current matrix elements. 
In this sense, $\psi_S$ and $\psi_D$ 
are associated with distributions of the \emph{total} momenta $\vec p$ 
and $\vec k$ of source and detector particle, respectively.

In appendix~\ref{integrations}, 
the amplitude $\ma$ at a stage before integrations are performed is 
displayed in equation~(\ref{AA}).

Since in the following we will not to consider the weak matrix elements and  
all final states are treated on the same footing as plane waves, we are 
allowed to use a simplified notation where
\begin{equation}
p'_S = \left( \begin{array}{c} E'_S \\[1mm] {\vec p}_S^{\,\prime} 
\end{array} \right)
\quad \mbox{and} \quad 
k'_D = \left( \begin{array}{c} E'_D \\[1mm] {\vec k}_D^{\,\prime} 
\end{array} \right)
\end{equation}
denote the \emph{sum} over all final momenta in the source and detection 
process, respectively. We furthermore denote the 4-momentum of the neutrino 
travelling from $\vec x_S$ to $\vec x_D$ by $q$. This is the momentum that 
appears in the neutrino propagator.

The following integrations have to be performed in order to obtain the 
oscillation amplitude $\ma$~\cite{grimus2019}:
\begin{enumerate}
\item 
$\int \dd^3 x_1$ and $\int \dd^3 x_2$,
\item
$\int \dd^3p$ and $\int \dd^3k$,
\item
$\int \dd t_1$ and $\int \dd t_2$,
\item
$\int \dd^3 q$ in the asymptotic limit 
$L \equiv \left| \vec x_D - \vec x_S \right| \to \infty$,
\item
$\int \dd q^0$ in an approximation that will be discussed in section~\ref{Q}.
\end{enumerate}
The first three integrations are done in appendix~\ref{integrations}.

Implicitly, in the time evolution in equations~(\ref{psiS}) and~(\ref{psiD}), 
we have made another assumption without which 
it would be difficult to write down an oscillation amplitude.
\begin{quote}
\refstepcounter{assumption}\label{env}
{\bf Assumption~\arabic{assumption}:} 
The source particle behaves approximately as a free particle in the time 
interval of length $\Delta t$ between two collisions. The same is assumed 
for the detection particle during the measurement time-interval of 
length $\tau$. 
\end{quote}
In other words, we assume that for short time intervals 
the time evolution of the source and detector particle can be written as 
a superposition of plane waves as done 
in equations~(\ref{psiS}) and~(\ref{psiD}), respevtively.
This is presumably a good approximation in the case of accelerator 
neutrinos produced by charged pions in a decay tunnel. 
It also plausible that, in the case of reactor neutrinos, the $\beta$-decay of 
a fission product is not influenced by collisions because nuclei are 
protected by an electron shell; nevertheless nuclei get kicked around by the 
environment and $\Delta t$ will be very short in this case. Maybe 
the environment can be partly taken into account by 
choosing suitable dispersion relations $E_S(\vec p\,)$ and 
$E_D(\vec k\,)$ or by allowing for random changes of 
$\vec \beta_S$ and $\vec \beta_D$.

We depart from the amplitude of equation~(\ref{A0}), 
taking into account equations~(\ref{subs}), (\ref{bw}) and~(\ref{ttheta}), 
in order to discuss the remaining integrations.  
Through the integration over $\vec q$ in the asymptotic limit $L \to \infty$, 
for each neutrino species with mass $m_j$, 
the neutrino 4-momentum $q$ becomes on-shell, the denominator of the 
neutrino propagator is cancelled and the amplitude gets a factor 
$L^{-1}$~\cite{GS}.\footnote{The separation between source and detection 
in the asymptotic limit $L \to \infty$ is a special case of 
localization effects in QFT that have recently been discussed 
in~\cite{karamitros}.} 
Moreover, the spatial neutrino momenta point 
from $\vec x_S$ to $\vec x_D$. Thus we arrive at the neutrino 4-momenta
\begin{equation}\label{q_j}
q_j = \left( \begin{array}{c} Q \\ \vec q_j 
\end{array} \right)
\quad \mbox{with} \quad 
\vec q_j = \sqrt{Q^2 - m_j^2}\, \vec \ell, \quad \vec\ell = 
\frac{\displaystyle \vec x_D - \vec x_S}{\displaystyle L}
\quad \mbox{and} \quad \left|\vec \ell\,\right| = 1.
\end{equation}
Defining~\cite{GSM} 
\begin{equation}
E_{Sj} = \left. E_S \right|_{\vec p = {\vec p}^{\,\prime}_S + \vec q_j},
\quad
E_{Dj} = \left. E_D \right|_{\vec k = {\vec k}^{\,\prime}_D - \vec q_j},
\end{equation}
the oscillation amplitude now reads
\begin{subequations}\label{A}
\begin{eqnarray}
\ma & \propto & \frac{1}{L} \sum_j U_{\beta j} U_{\alpha j}^* \int \dd Q
\label{AU} \\ &&
\times \psi_S \left( {\vec p}^{\,\prime}_S + \vec q_j \right)
\,\frac{1 - \exp \left[ \left( -i \left(E_{Sj} - E'_S - Q \right) - 
\frac{\displaystyle \Gamma m_S}{\displaystyle 2E_{Sj}} 
\right) \Delta t \right]}{i \left(E_{Sj} - E'_S - Q \right) +  
\frac{\displaystyle \Gamma m_S}{\displaystyle 2E_{Sj}}}
\label{AS} \\ &&
\times \exp \left( -i \left( t_D - t_S \right)Q + iL \sqrt{Q^2 - m_j^2}
\,\right)
\label{AO} \\ &&
\times \psi_D \left( {\vec k}^{\,\prime}_D - \vec q_j \right)
\widetilde \Theta_\tau \left( E_{Dj} - E'_D + Q \right). 
\label{AD}
\end{eqnarray}
\end{subequations}
The further discussion is based on this amplitude.

The amplitude in equation~(\ref{A}) has been derived for neutrinos. How is 
this amplitude modified for antineutrinos? 
In this case, the neutrino propagator 
in equation~(\ref{A0}) has $x_2 - x_1$ instead of $x_1 - x_2$ and 
the complex conjugate is on the CKM matrix element 
$U_{\beta j}$ instead of $U_{\alpha j}$. 
Changing the neutrino momentum $q$ to $q' = -q$, we are back to 
equation~(\ref{A}), apart from potential 
effects in the matrix elements which do not concern us here and  
the shift of the complex conjugation from $U_{\alpha j}$ to $U_{\beta j}$. 
In this way, the further discussion 
can be trivially modified for antineutrino oscillations.

Now we turn to the question whether $\psi_S$ and $\psi_D$ 
can induce decoherence in the oscillation amplitude. 
Let us for instance consider $\psi_S$. 
Suppose the width of this wave packet is $\sigma_S$ and we have two 
massive neutrino species with masses $m_i$ and $m_j$ ($m_i \neq m_j$). 
In order to have oscillations involving these two massive neutrinos, neither 
should be suppressed by $\psi_S$. Therefore, we simultaneously require
\begin{equation}
\left| {\vec p}^{\,\prime}_S + \vec q_i \right| \lesssim \sigma_S,
\quad
\left| {\vec p}^{\,\prime}_S + \vec q_j \right| \lesssim \sigma_S.
\end{equation}
Consequently, using the triangle inequality and
expanding the square root in equation~(\ref{q_j}), we obtain
\begin{equation}\label{sigma}
\left| \vec q_i - \vec q_j \right| \simeq
\left| \frac{m_i^2 - m_j^2}{2 \bar Q} \right| \lesssim 2\sigma_S.
\end{equation}
In this equation, $\bar Q$ is some average neutrino energy. 
In section~\ref{appr} we will specify $\bar Q$.
Denoting by $\Delta m^2$ some generic and positive 
neutrino mass-squared difference and 
assuming that via the uncertainty relation the width $\sigma_{xS}$ of 
$\psi_D$ in coordinate space is approximately given by 
$\sigma_{xS} \sim 1/(2\sigma_S)$, we reformulate equation~(\ref{sigma}) as
\begin{equation}
\frac{2 \bar Q}{\Delta m^2} \gtrsim \sigma_{xS}
\quad \mbox{or} \quad 
L_\mathrm{osc} \equiv 4\pi\, \frac{\bar Q}{\Delta m^2} \gtrsim 
2\pi\, \sigma_{xS}.
\end{equation}
Since the oscillation length $L_\mathrm{osc}$ is a macroscopic quantity 
whereas $\sigma_{xS}$ is microscopic, we conclude that 
$\psi_S$ does not cause any decoherence effect. By analogy, the same 
is true for $\psi_D$~\cite{G2003,GSM}.

It remains to investigate potential decoherence caused by the second factor 
in equation~(\ref{AS}) and by $\widetilde\Theta_\tau$ in equation~(\ref{AD}). 
In this context, we use henceforth 
the specific form of $\Theta_\tau$ of equation~(\ref{Thetatau}). Therefore,
\begin{equation}\label{ThetaE}
\widetilde\Theta_\tau(E) = 2\,\frac{\sin (E \tau/2)}{E}. 
\end{equation}

\boldmath
\section{Integration over the neutrino energy $Q$}
\unboldmath
\label{Q}
The integration over $Q$ cannot be performed exactly. However, since it 
is essentially limited by $\widetilde\Theta_\tau$ in equation~(\ref{AD}), 
we can make reasonable approximations. We read off from 
equation~(\ref{ThetaE}) that the first zeros of this function are 
$E = \pm \pi/\tau$. We will take this as the typical energy scale associated 
with $\Theta_\tau$. However, since this is a slowly decreasing function, 
we might for instance integrate to the 100th zero, 
\textit{i.e.}\ over the range of $Q$ defined by 
$\left| E_{Dj} - E'_D + Q \right| \lesssim 100 \pi/\tau$, 
to cover the bulk of the $Q$-integral.

\subsection{Assumptions and approximations}
\label{appr}
In realistic cases, $\pi/\tau$ is many orders of magnitude 
below 1\,MeV while the typical energy scale occurring in the matrix elements 
at neutrino source and detection is 1\,MeV or larger. Taking into account 
the approximate energy range of $Q$ discussed above, we are lead to
\begin{quote}
\refstepcounter{assumption}\label{ma}
{\bf Assumption~\arabic{assumption}:} The $Q$-dependence of the 
weak-current matrix elements in the source and the detection process 
can be neglected in the $Q$-integration. 
\end{quote}
Another limitation of the $Q$-integration which can be read off from 
equation~(\ref{q_j}) is 
\begin{equation}
Q > \max_j m_j.
\end{equation}
However, this limitation is irrelevant because 
the relevant neutrino energies in the integration fulfill $Q \ggg m_j$.

In the following, the strategy will be to define a suitable mean neutrino 
energy $\bar Q$ and a new integration variable $\delta Q$ via 
\begin{equation}
Q = \bar Q + \delta Q.
\end{equation}
We will expand all relevant quantities that depend on $Q$ in the small 
quantities $\delta Q$ and $m_j^2/(2\bar Q)$. In particular, it is 
important to expand the square root in the 
neutrino momenta $\vec q_j$, equation~(\ref{q_j}), for performing the 
integration over $\delta Q$. Making an expansion in $1/\bar Q$, we find
\begin{equation}\label{sqrtQ}
\sqrt{Q^2 - m_j^2} = \bar Q + \delta Q - \frac{m_j^2}{2 \bar Q} +
\frac{m_j^2 \delta Q}{2 {\bar Q}^2} + \cdots,
\end{equation}
where the dots indicate terms of order $1/{\bar Q}^3$ and higher.
We will content ourselves with the 
first three terms on the right-hand side of this equation, except in 
section~\ref{time} where we also discuss the fourth term.

We define a mean neutrino energy $\bar Q$ via the detection process by 
requiring exact energy conservation in the approximation of vanishing 
neutrino masses. This means that $\bar Q$ is the solution of 
\begin{equation}\label{DQbar}
\left. E_D \right|_{\vec k = {\vec k}^{\,\prime}_D - \bar Q \vec\ell} 
 = E'_D - \bar Q.
\end{equation}
This allows us to define approximate energies~\cite{grimus2019}
\begin{equation}
\bar E_S = \left. E_S \right|_{\vec p = {\vec p}^{\,\prime}_S + \bar Q \vec\ell}
\quad \mbox{and} \quad
\bar E_D = \left. E_D \right|_{\vec k = {\vec k}^{\,\prime}_D - \bar Q \vec\ell}
\end{equation}
of source and detector particle, respectively. 
Note that because of equation~(\ref{DQbar}) the relation
\begin{equation}\label{EDQ}
\bar E_D + \bar Q - E'_D = 0
\end{equation}
is valid. The relation corresponding to the source process is, however, 
non-zero in general since $\bar Q$ is defined via the detection process. 
We rather have
\begin{equation}
\bar E_S - \bar Q - E'_S = \bar E_S + \bar E_D - E'_S - E'_D
\equiv \Delta \bar E,
\end{equation}
where $\Delta\bar E$ is a kind of measure of energy non-conservation 
in the compound source--detection process. 
We emphasize that here and in the following all barred quantities 
refer to the mean energy $\bar Q$ defined in equation~(\ref{DQbar}).

Now, based on equation~(\ref{sqrtQ}), 
we include $\delta Q$ and $m_j^2/(2 \bar Q)$ in first order 
in the further computations. In this way we obtain
\begin{subequations}
\begin{eqnarray}
\label{ESj}
E_{Sj} = \bar E_S + \bar\beta_S \left( \delta Q - \frac{m_j^2}{2 \bar Q} 
\right)
& \mbox{with} &
\bar\beta_S = \left. \vec\ell \cdot \left(\vec \nabla_p E_S 
\right)\right|_{\vec p = {\vec p}^{\,\prime}_S + \bar Q \vec\ell},
\\
\label{EDj}
E_{Dj} = \bar E_D - \bar\beta_D \left( \delta Q - \frac{m_j^2}{2 \bar Q} 
\right)
& \mbox{with} &
\bar\beta_D = \left. \vec\ell \cdot \left(\vec \nabla_k E_D 
\right)\right|_{\vec k = {\vec k}^{\,\prime}_D - \bar Q \vec\ell}.
\end{eqnarray}
\end{subequations}
In these equations, $\bar\beta_S$ and $\bar\beta_D$ are mean velocities of 
the source and detector particle, respectively, projected 
unto the direction $\vec\ell$. The expressions we eventually need are
\begin{subequations}
\begin{eqnarray}
E_{Sj} - Q - E'_S &=& - \left( 1 - \bar \beta_S \right) \delta Q -
\bar\beta_S\, \frac{m_j^2}{2 \bar Q} + \Delta \bar E,
\\
E_{Dj} + Q - E'_D &=& \hphantom{-}
\left( 1 - \bar \beta_D \right) \delta Q + 
\bar\beta_D\, \frac{m_j^2}{2 \bar Q}.
\end{eqnarray}
\end{subequations}

In order to integrate over $\delta Q$ we finally resort to
\begin{quote}
\refstepcounter{assumption}\label{infty}
{\bf Assumption~\arabic{assumption}:} We assume that, in the integration 
over $\delta Q$, we are allowed to extend the integration limits to 
minus and plus infinity. In addition, we neglect the $\delta Q$-dependence 
in the wave functions $\psi_S$ and $\psi_D$. 
\end{quote}
The latter implies that the widths $\sigma_S$ and $\sigma_D$ corresponding 
to $\psi_S$ and $\psi_D$, respectively, are large enough such 
that this neglect is a good approximation, \textit{i.e.}
\begin{equation}\label{sigmaSD}
\sigma_S \gg \frac{\pi}{\tau}, \quad 
\sigma_D \gg \frac{\pi}{\tau}.
\end{equation}

At least for the detector particle the inequality is easy to check. 
Since the detector particle is at rest apart from thermal motion, 
we estimate~\cite{grimus2019}
\begin{equation}
\sigma_D \sim \sqrt{3 m_D k_BT}.
\end{equation}
For instance, for a proton as detector particle we find 
$\sigma_D \sim 0.01$\,MeV at room temperature. Let us assume 
$\sigma_S \sim \sigma_D$, $\tau \sim 10^{-12}$\,s, $m_j \lesssim 0.1$\,eV and
$\bar Q \lesssim 0.5$\,MeV. Then
\begin{equation}
\frac{\pi}{\tau} \sim 2 \times 10^{-9}\,\mathrm{MeV}, \quad
\frac{m_j^2}{2\bar Q} \lesssim 10^{-14}\,\mathrm{MeV}
\end{equation}
and these quantities are much smaller than the widths $\sigma_S$ 
and $\sigma_D$. 

In summary, we have argued that, in the range of $\delta Q$ where 
$\Theta_\tau$ is unsuppressed, we are allowed to neglect the 
$\delta Q$-dependence of the weak matrix elements and of $\psi_S$ and 
$\psi_D$ while, in the range of $\delta Q$ where this dependence 
becomes relevant, $\Theta_\tau$ is suppressed. Therefore, it is a reasonable 
approximation to integrate over $\delta Q$ from $-\infty$ to $+\infty$ 
with constant neutrino energy $\bar Q$ in the weak matrix elements and in 
$\psi_S$ and $\psi_D$.

\subsection{Integration over \boldmath$\delta Q$\unboldmath}
In order to apply the integral formula of appendix~\ref{integral}, 
we define the times
\begin{equation}\label{lt}
\lambda = \left( 1 - \bar\beta_D \right) \frac{\tau}{2},
\quad
t_c = t_D - t_S - L
\end{equation}
and the quantities
\begin{subequations}\label{z}
\begin{eqnarray}
z_{1j} &=& -\frac{1}{1 - \bar\beta_S} \left( 
\frac{\bar\beta_S m_j^2}{2\bar Q} + i \frac{\Gamma m_S}{2\bar E_S} 
- \Delta \bar E \right),
\label{z1}
\\
z_{2j} &=& -\frac{1}{1 - \bar\beta_D} \, \frac{\bar\beta_D m_j^2}{2\bar Q}.
\label{z2}
\end{eqnarray}
\end{subequations}
Note that in $\Gamma m_S/E_{Sj}$ we have replaced $E_{Sj}$ by $\bar E_S$ 
since $\Gamma$ is already pretty small, \textit{cf.}\ section~\ref{function}.
Furthermore, we have to consider the exponential function, 
equation~(\ref{AO}), 
and expand the exponent according to equation~(\ref{sqrtQ}):
\begin{equation}\label{exp}
\exp \left( -i (t_D - t_S) Q + i L \sqrt{Q^2 - m_j^2} \right) = 
\exp \left( -i t_c \bar Q - i\frac{m_j^2}{2\bar Q}\,L \right) \times
\exp \left( -it_c\, \delta Q \right).
\end{equation}
The first factor on the right-hand side contains the irrelevant phase 
$t_c \bar Q$, but also the phase that gives rise to neutrino oscillations. 
The second factor has to be taken into account in the integral. 
With assumption~\ref{infty} we rewrite equation~(\ref{A}) as
\begin{subequations}\label{A4}
\begin{eqnarray}
\ma & \propto & \frac{1}{L}\, 
\psi_S \left( {\vec p}^{\,\prime}_S + \bar Q \vec\ell\, \right)
\psi_D \left( {\vec k}^{\,\prime}_D - \bar Q \vec\ell\, \right) 
e^{-i t_c\bar Q}\, \sum_j U_{\beta j} U_{\alpha j}^*\, e^{-i m_j^2 L/(2 \bar Q)} 
\\ && \times \label{A4I}
\frac{\tau}{2} \, \frac{1}{1- \bar\beta_S} \left[
I_j(t_c) - I_j(t_c - T)\, e^{-iT z_{1j}} \right]
\end{eqnarray}
\end{subequations}
with
\begin{equation}\label{T}
T \equiv (1- \bar\beta_S) \Delta t.
\end{equation}
Note that $T$ is positive. 
The integrals in equation~(\ref{A4I}) have the form 
\begin{equation}\label{Ij}
I_j(t) =  2i \int \dd \delta Q\, 
\frac{e^{-it\, \delta Q}}{\delta Q - z_{1j}} 
\,\frac{\sin \left(\lambda ( \delta Q - z_{2j}) \right)}%
{\lambda ( \delta Q - z_{2j})}
\end{equation}
with $t = t_c$ and $t = t_c - T$. They are thus of the type of 
equation~(\ref{Ix}) treated 
in appendix~\ref{integral} and the results of this appendix can be applied.

It is useful to define 
\begin{equation}\label{M}
M_j\left(t_c,\Delta\bar E\right) = 
\frac{\tau}{2} \, \frac{1}{1- \bar\beta_S} \left[
I_j(t_c) - I_j(t_c - T)\, e^{-iT z_{1j}} \right].
\end{equation}
This corresponds to equation~(\ref{A4I}). 
Note that the result of the integral $I(c)$, 
equation~(\ref{I}), requires three case distinctions. Since 
in $M_j\left(t_c,\Delta\bar E\right)$ there are two such integrals, 
this quantity requires five case distinctions, depending on the 
relative positions of the intervals 
\begin{equation}\label{intervals}
\mathcal{I}_c = (t_c-T,t_c) \quad \mbox{and} \quad
\mathcal{I}_\lambda = (-\lambda,\lambda). 
\end{equation}
The two cases with 
$\mathcal{I}_c \cap \mathcal{I}_\lambda = \emptyset$ simply yield 
$M_j\left(t_c,\Delta\bar E\right) = 0$.
This is trivial for $t_c < -\lambda$, \textit{cf.}\ equation~(\ref{Ia}), but  
for $\lambda < t_c - T$ equation~(\ref{Ic}) has to be invoked. 
Among the remaining three cases with 
$\mathcal{I}_c \cap \mathcal{I}_\lambda \neq \emptyset$ there is one case 
where the intervals completely overlap. In this situation there are 
two possibilities: 
\begin{itemize}
\item
$T < 2\lambda$ with $\mathcal{I}_c \subset \mathcal{I}_\lambda$ and
\item
$T > 2\lambda$ with $\mathcal{I}_\lambda \subset \mathcal{I}_c$.
\end{itemize}
For the sake of clarity we present the result for both possibilities 
separately in spite of some overlap:
\begin{subequations}\label{Mres1}
\begin{align}
\fbox{$T < 2\lambda$:} &&&& \nonumber \\
-\infty <\, &t_c < -\lambda & &\Rightarrow & 
M_j/f &= 0,
\\
-\lambda <\, &t_c < -\lambda + T & &\Rightarrow & 
M_j/f &=
\frac{e^{-it_cz_{2j}}}{z_{1j} - z_{2j}} 
\left( 1 - e^{-i(\lambda + t_c)\left(z_{1j} - z_{2j}\right)} \right),
\\
-\lambda + T <\, &t_c < \lambda & &\Rightarrow & 
M_j/f &=
\frac{e^{-it_cz_{2j}}}{z_{1j} - z_{2j}} 
\left( 1 - e^{-iT\left(z_{1j} - z_{2j}\right)} \right),
\\
\lambda <\, &t_c < \lambda + T & &\Rightarrow & 
M_j/f &=
\frac{e^{-it_cz_{2j}}}{z_{1j} - z_{2j}} 
\left( e^{i(\lambda - t_c) \left(z_{1j} - z_{2j}\right)} - 
e^{-iT\left(z_{1j} - z_{2j}\right)} \right),
\\
\lambda + T<\, &t_c < \infty & &\Rightarrow & 
M_j/f &= 0,
\end{align}
\end{subequations}
\begin{subequations}\label{Mres2}
\begin{align}
\fbox{$T > 2\lambda$:} &&&& \nonumber \\
-\infty <\, &t_c < -\lambda & &\Rightarrow & 
M_j/f &= 0,
\\
-\lambda <\, &t_c < \lambda & &\Rightarrow & 
M_j/f &=
\frac{e^{-it_cz_{2j}}}{z_{1j} - z_{2j}} 
\left( 1 - e^{-i(\lambda + t_c)\left(z_{1j} - z_{2j}\right)} \right),
\\
\lambda <\, &t_c < -\lambda + T & &\Rightarrow & 
M_j/f &=
\frac{e^{-it_cz_{2j}}}{z_{1j} - z_{2j}} 
\left( e^{i(\lambda - t_c)\left(z_{1j} - z_{2j}\right)} - 
e^{-i(\lambda + t_c) \left(z_{1j} - z_{2j}\right)} \right),
\\
-\lambda + T <\, &t_c < \lambda + T & &\Rightarrow & 
M_j/f &=
\frac{e^{-it_cz_{2j}}}{z_{1j} - z_{2j}} 
\left( e^{i(\lambda - t_c) \left(z_{1j} - z_{2j}\right)} - 
e^{-iT\left(z_{1j} - z_{2j}\right)} \right),
\\
\lambda + T<\, &t_c < \infty & &\Rightarrow & 
M_j/f &= 0.
\end{align}
\end{subequations}
There is a common factor 
\begin{equation}
f = -\frac{2\pi i}{\left( 1-\bar\beta_S \right)\left( 1-\bar\beta_D \right)}
\end{equation}
in all cases. For the sake of brevity we have omitted in 
equations~(\ref{Mres1}) and~(\ref{Mres2}) 
the dependence on the variables $t_c$ and $\Delta\bar E$ in $M_j$. 
It is easy to check that $M_j\left(t_c,\Delta\bar E \right)$ 
is continuous in $t_c$ for both $T < 2\lambda$ and $T > 2\lambda$. 
This is necessarily so because $I(c)$ in equation~(\ref{I}) is continuous.

\section{Time correlation between neutrino production and detection}
\label{time}
Equations~(\ref{Mres1}) and~(\ref{Mres2}) lead to a correlation between 
the times $t_S$ and $t_D$~\cite{giunti1993} because a non-zero 
$M_j\left(t_c,\Delta\bar E\right)$ is only possible for 
\begin{equation}\label{ineq}
-\lambda < t_c < \lambda + T.
\end{equation}
Usually it is not known when the source particle and the neutrino are 
produced, but the 
time $t_D \pm \tau/2$ of neutrino detection is recorded. Therefore, 
equation~(\ref{ineq}) allows to infer the time $t_S$ when the 
source particle is created:
\begin{equation}\label{t_S}
t_D - \lambda - L - T < t_S < t_D + \lambda - L.
\end{equation}
This is an automatic consequence of the formalism and 
lends credibility to our neutrino oscillation model~\cite{grimus2019}. 

Let us compare equation~(\ref{t_S}) with a purely classical consideration. We 
keep the measurement interval $t_D \pm \tau/2$ fixed and determine the 
earliest production time $t_{S1}$ and the latest 
production time $t_{S2}$ of the source particle 
such that a neutrino measurement is possible. Therefore, $t_S$ lies in 
the interval 
\begin{equation}
t_{S1}< t_S < t_{S2}.
\end{equation}
\paragraph{Determination of \boldmath$t_{S1}$\unboldmath:}
In this case the source particle decays at the latest possible 
time $t_{S1} + \Delta t$ at the point $x_{S1} = x_S + \bar\beta_S \Delta t$ 
and the neutrino arrives at the time 
$t_D - \tau/2$, which is the earliest possible measurement time, at 
the point $x_{D1} = x_D - \bar\beta_D \tau/2$.\footnote{We 
assume that the neutrino travels with the speed of light.} 
Therefore, the condition for the determination of $t_{S1}$ is given by
\begin{equation}
x_{D1} = x_{S1} + t_D - \frac{\tau}{2} - t_{S1} - \Delta t 
\quad \Rightarrow \quad
t_{S1} = t_D - \lambda - L - T.
\end{equation}
\paragraph{Determination of \boldmath$t_{S2}$\unboldmath:}
Now the source particle decays at the time $t_{S2}$ at the 
point $x_{S2} = x_S$ and the neutrino arrives at the time 
$t_D + \tau/2$ at the point $x_{D2} = x_D + \bar\beta_D \tau/2$. 
This leads to 
\begin{equation}
x_{D2} = x_{S2} + t_D + \frac{\tau}{2} - t_{S2} 
\quad \Rightarrow \quad
t_{S2} = t_D + \lambda - L.
\end{equation}
In this way, taking into account 
positions and velocities of the source and detector particle and the time of 
flight of the neutrino, we obtain full agreement with equation~(\ref{t_S}) 
derived from a QFT formalism.

It is interesting that even an analogue to the 
``separation of neutrino wave packets''~\cite{giunti1993,kayser} 
is contained in our model.
To understand this point we have to take a look at equation~(\ref{sqrtQ}). 
Up to now we have only taken into account the first three terms on the 
right-hand side of this equation. However, it is the fourth term that 
leads to this effect~\cite{grimus2019}. If we take it into account, the 
quantity $t_c$ depends on the index~$j$:
\begin{equation}
t_{cj} = t_D -t_S - L - L \,\frac{m_j^2}{2{\bar Q}^2}.
\end{equation}
How do we have to interpret of the last term? 
The velocity of the massive neutrino $\nu_j$ with energy $\bar Q$ is 
\begin{equation}
\beta_j = \frac{\sqrt{{\bar Q}^2 - m_j^2}}{\bar Q} \simeq 
1 - \frac{m_j^2}{2{\bar Q}^2}.
\end{equation}
Therefore, it covers the distance $L$ in the time
\begin{equation}
\frac{L}{\beta_j} \simeq L \left( 1 + \frac{m_j^2}{2{\bar Q}^2} \right) 
\equiv L + \delta t_j.
\end{equation}
Consequently, $\delta t_j$ is the time delay due to the finite neutrino 
mass. It has to be incorporated into equation~(\ref{ineq}): 
\begin{equation}\label{ineq_j}
-\lambda < t_{cj} < \lambda + T.
\end{equation}

If we have two neutrino masses $m_i$ and $m_j$ ($m_i \neq m_j$), both 
$t_{ci}$ and $t_{cj}$ have to fulfill equation~(\ref{ineq_j}) in order to 
guarantee coherence.  
Assuming $m_i < m_j$ which in turn implies $t_{ci} > t_{cj}$, coherence 
between $\nu_i$ and $\nu_j$ is lost for
\begin{equation}
t_{ci} \in (-\lambda, \lambda + T), \;\; t_{cj} < -\lambda
\quad \mbox{or} \quad
t_{cj} \in (-\lambda, \lambda + T), \;\; t_{ci} > \lambda + T.
\end{equation}
In practice, however, $\delta t_j$ is so small that this decoherence 
effect is irrelevant. For instance, setting
$m_j = 0.1$\,eV, $\bar Q = 0.5$\,MeV, $L = 12000$\,km, we find 
$t_j = 0.8 \times 10^{-15}$\,s. This is roughly the maximal possible 
time delay $\delta t_j$ for (future) neutrino oscillation experiments 
on earth.

\section{The function \boldmath$|M_j\left(t_c,\Delta\bar E\right)|^2$
\unboldmath}
\label{function}
In typical neutrino-oscillation experiments the decay width $\Gamma$ is 
very small. For instance, 
mean lives of fission products in a reactor are 
rather large. Assuming that the bulk of reactor neutrinos comes from 
mean lives larger than 1\,second, we obtain extremely small decay widths:
\begin{equation}
\Gamma \lesssim \frac{\hbar}{1\,\mathrm{s}} \simeq 6.6 \times 10^{-22}\,
\mathrm{MeV}.
\end{equation}
Accelerator neutrinos are produced by the decay of charged pions in which case 
\begin{equation}
\Gamma \simeq 2.5 \times 10^{-14}\,\mathrm{MeV}.
\end{equation}
Since the pions are relativistic, the effective width 
$\Gamma m_S/\bar E_S$ is even smaller by several orders of magnitude.

The smallness of $\Gamma$ in the denominator in equation~(\ref{AS}) 
suggests a potential source of decoherence. Indeed, inspecting 
equation~(\ref{z}) we are tempted deduce a coherence condition
\begin{equation}
\left| \left( \bar\beta_S - \bar\beta_D \right) 
\frac{\Delta m^2}{2\bar Q} \right| \lesssim \frac{\Gamma m_S}{\bar E_S}.
\end{equation}
This inequality could be violated even for thermal velocities 
$\bar\beta_S$, $\bar\beta_D$ and the small quantities 
\begin{equation}\label{small}
\frac{\Delta m^2_\mathrm{atm}}{2 \bar Q} \lesssim 2.5 \times 10^{-15}\,
\mathrm{MeV}
\quad \mbox{and} \quad 
\frac{\Delta m^2_\mathrm{sol}}{2 \bar Q} \lesssim 7 \times 10^{-17}\,
\mathrm{MeV}
\end{equation}
whose upper limits have been obtained from $\bar Q \gtrsim 0.5$\,MeV
and a three-neutrino fit~\cite{esteban,capozzi} to all available data
yielding
\begin{equation}
\Delta m^2_\mathrm{atm} \sim 2.5 \times 10^{-3}\,\mathrm{eV}^2
\quad \mbox{and} \quad 
\Delta m^2_\mathrm{sol} \sim 7 \times 10^{-5}\,\mathrm{eV}^2.
\end{equation}
However, the coherence condition above is fictitious. 

To prove this claim we consider the dependence 
on $\Delta \bar E$ of the function
\begin{equation}\label{Mj2}
\left| M_j\left(t_c,\Delta\bar E\right) \right|^2 = 
\frac{(2\pi)^2}{(1-\bar\beta_S)^2 (1-\bar\beta_D)^2}
\left| \frac{1 - 
e^{-i\mt(t_c) (z_{1j} - z_{2j})}}{z_{1j} - z_{2j}} \right|^2
\end{equation}
that occurs in the event rate or total cross section of the compound process.
With equations~(\ref{Mres1}) and~(\ref{Mres2}) 
it can easily be shown that
\begin{equation}\label{tc}
\mt(t_c) = \mathrm{length} 
\left( \mathcal{I}_c \cap \mathcal{I}_\lambda \right)
\end{equation}
for both $T < 2\lambda$ and $T > 2\lambda$.
The intervals $\mathcal{I}_c$ and $\mathcal{I}_\lambda$ are defined in 
equation~(\ref{intervals}).
Therefore, $\mt$ rises linearly from zero to $T$ if $T < 2\lambda$ 
or to $2\lambda$ if $T > 2\lambda$, then remains constant, and 
thereafter decreases linearly to zero.

Actually, if $t_c$ is fixed, 
$\left| M_j\left(t_c,\Delta\bar E\right) \right|^2$ 
is proportional to the probability distribution of $\Delta \bar E$.
Its maximum will give the preferred value of $\Delta \bar E$. It 
will approximately be reached at 
$\mathrm{Re} \left( z_{1j} - z_{2j} \right) = 0$ because $\Gamma$ is 
extremely small. According to equation~(\ref{z}), the corresponding
value of $\Delta \bar E$ depends on $m_j$. But whenever we choose 
a $t_c$ such that $\left| M_j\left(t_c,\Delta\bar E\right) \right|^2$ is 
safely away from zero, the order of magnitude of its width 
in the variable $\Delta \bar E$ is given by $1/T$ or $1/\tau$. 
Since it is reasonable to assume that both $1/T$ and $1/\tau$ are much 
larger than the quantities in equation~(\ref{small}), the peak of 
$\left| M_j\left(t_c,\Delta\bar E\right) \right|^2$ 
as a function of $\Delta \bar E$ is so broad that its maximum is 
practically independent of $m_j$.

So from this consideration we conclude that 
$M_j\left(t_c,\Delta\bar E\right)$ has nothing to do with 
coherence but it determines the goodness of total energy conservation, 
\textit{i.e.}\ how well $\Delta \bar E  = 0$ is fulfilled. Clearly, 
as discussed above, in terms of order of magnitude, 
energy conservation cannot be better than 
$1/\tau$ or $1/T$. In the further discussion we neglect all small 
quantities in $M_j\left(t_c,\Delta\bar E\right)$. 
In addition, for practical purposes we introduce
\begin{quote}
\refstepcounter{assumption}\label{rest}
{\bf Assumption~\arabic{assumption}:} We assume that, 
apart from thermal motion, the detector particle is at rest.
\end{quote}
Altogether we resort to the approximation
\begin{equation}\label{set}
\bar\beta_D = 0, \quad 
\bar E_D = m_D, \quad
z_{1j} - z_{2j} = \frac{\Delta \bar E}{1 - \bar\beta_S}, \quad 
\lambda = \frac{\tau}{2}.
\end{equation}
Note that we do not fix the value of $\bar\beta_S$ because we want the 
following discussion to be valid for both reactor and accelerator neutrinos. 
With the above approximations, $M_j\left(t_c,\Delta\bar E\right)$ 
is independent of $m_j$ and we drop the index $j$. In this way, we obtain
\begin{equation}\label{MM}
\left| M\left(t_c,\Delta\bar E\right) \right|^2 = (2\pi)^2 \times
\left| \frac{1 - 
e^{-i\mt(t_c) \Delta \bar E/(1-\bar\beta_S)}}{\Delta \bar E} \right|^2.
\end{equation}

When a neutrino event is recorded in the detector, it is unknown when the 
corresponding source particle has been produced. Suppose source 
particles are created at constant rate $N_S$. Then in a 
time interval $\dd t_S$ the number of source particles created is 
$N_S \dd t_S$. Therefore, in order to compute the event rate in the detector, 
one has to integrate over $t_S$ or, equivalently, over $t_c$. 
Thus we have to compute the integral 
\begin{equation}
\int_{-\lambda}^{\lambda + T} \dd t_c 
\left| M\left(t_c,\Delta\bar E\right) \right|^2 = 
4 \times (2\pi)^2 \int_{-\lambda}^{\lambda + T} \dd t_c 
\left( \frac{\sin \left[\mt(t_c) \Delta \bar E/\left( 2(1-\bar\beta_s) \right)
\right]}{\Delta \bar E} \right)^2.
\end{equation}
In order to present the result of the integration we define the functions
\begin{equation}
\delta_1(E,t) = t f_1(Et)
\quad \mbox{and} \quad
\delta_2(E,t) = t f_2(Et)
\end{equation}
with
\begin{equation}
f_1(y) = \frac{1}{\pi} \left( \frac{\sin y}{y} \right)^2
\quad \mbox{and} \quad
f_2(y) = \frac{2}{\pi} \left( \frac{1}{y^2} - \frac{\sin y}{y^3} \right),
\end{equation}
respectively. Note that 
\begin{equation}
\int_{-\infty}^\infty \dd y \, f_1(y) = 
\int_{-\infty}^\infty \dd y \, f_2(y) = 1,
\end{equation}
whence it follows that
\begin{equation}
\lim_{t \to \infty} \delta_1(E,t) = \lim_{t \to \infty} \delta_2(E,t) = 
\delta(E).
\end{equation}
The integration yields
\begin{equation}\label{M21}
\int_{-\lambda}^{\lambda + T} \dd t_c 
\left| M\left(t_c,\Delta\bar E\right) \right|^2 = (2\pi)^3   
\Delta t \left[ \left( \tau - T \right) \delta_1(\Delta\bar E, \Delta t/2) + 
T \delta_2(\Delta\bar E, \Delta t) \right]
\quad \mbox{for} \;\; T < \tau
\end{equation}
and
\begin{eqnarray}
\lefteqn{
\int_{-\lambda}^{\lambda + T} \dd t_c 
\left| M\left(t_c,\Delta\bar E\right) \right|^2} 
\nonumber \\ && \label{M22}
= (2\pi)^3   
\frac{\tau}{1 - \bar\beta_S}  
\left[ \left( T - \tau \right) 
\delta_1 \left( \Delta\bar E, \frac{\tau}{2(1 - \bar\beta_S)} \right) + 
\tau\, \delta_2\left( \Delta\bar E, \frac{\tau}{1 - \bar\beta_S} \right) 
\right]
\\ \nonumber &&
\hfill \mbox{for} \quad T > \tau.
\end{eqnarray}

We expect that for all practical purposes the approximation 
\begin{equation}
\delta_1(\Delta\bar E,t/2) \simeq \delta_2(\Delta\bar E,t) \simeq 
\delta(\Delta\bar E)
\quad \mbox{for} \quad 
t = \Delta t, \quad \frac{\tau}{1 - \bar\beta_S}
\end{equation}
holds. Consequently, both equations~(\ref{M21}) and~(\ref{M22}) 
approximately yield
\begin{equation}
\int_{-\lambda}^{\lambda + T} \dd t_c 
\left| M\left(t_c,\Delta\bar E\right) \right|^2 \simeq 
(2\pi)^3  \tau \Delta t\,\delta(\Delta\bar E) 
\end{equation}
with a $\delta$-function $\delta(\Delta\bar E)$. Alternatively we may define
\begin{equation}
\int_{-\lambda}^{\lambda + T} \dd t_c 
\left| M\left(t_c,\Delta\bar E\right) \right|^2 = 
(2\pi)^3  \tau \Delta t\,\bar\delta(\Delta\bar E),
\end{equation}
where the definition of $\bar\delta(\Delta\bar E)$ can be read off from 
equations~(\ref{M21}) and~(\ref{M22}). In $\bar\delta(\Delta\bar E)$ we have 
suppressed the dependence on $\tau$, $\Delta t$ and $\bar\beta_S$ that 
disappears in the limit $\bar\delta(\Delta\bar E) \to \delta(\Delta\bar E)$.

\section{Reactor and accelerator neutrinos}
\label{accelerator}
In addition to equation~(\ref{set}), for reactor neutrinos a reasonable 
approximation is 
\begin{equation}
\bar\beta_S = 0, \quad T = \Delta t, \quad \bar E_S = m_S,
\end{equation}
while for accelerator neutrinos, where 
the typical source particle is a relativistic charged pion and $m_S = m_\pi$, 
we have 
\begin{equation}\label{acc}
1 - \bar \beta_S \simeq 
\frac{\displaystyle m_S^2}{\rule{0pt}{12pt}\displaystyle 2\bar E_S^2}, 
\quad
T \simeq 
\frac{\displaystyle m_S^2}{\rule{0pt}{12pt}\displaystyle 2\bar E_S^2}\,
\Delta t.
\end{equation}

In order to get feeling for the orders of magnitude, we choose some reasonable 
numbers for accelerator neutrinos: 
$m_S = m_\pi \simeq 140$\,MeV, $\bar E_S = 5$\,GeV, and for the length 
of the decay tunnel we assume 300\,m. Since the pion is relativistic, with 
this length of the decay tunnel we obtain $\Delta t \simeq 10^{-6}$\,s.
Therefore, using equation~(\ref{acc}), we find 
$T \simeq 4 \times 10^{-10}$\,s. 
Moreover, with $\Gamma \simeq 2.5 \times 10^{-14}$\,MeV we obtain  
$\Gamma m_S/\bar E_S \simeq 0.7 \times 10^{-15}$\,MeV.

Let us collect all terms in the event rate of the compound process 
we have discussed and include the factors that we have left out so far:
\begin{subequations}\label{R}
\begin{eqnarray}
&& \frac{1}{(2\pi)^3\, 2 \bar E_S} \times (2\pi)^6 \left| \psi_S \right|^2 
\left| \mathcal{M}_S \right|^2 
\prod_{i=1}^{n_S-1} \frac{\dd^3 p'_i}{(2\pi)^3 2 E'_{pi}}
\label{Ra} \\ && \times
\frac{1}{(2\pi)^8} \left( \frac{2\pi^2}{L} \right)^2 P_{\nu_\alpha\to\nu_\beta}
\label{Rb} \\ && \times
\frac{1}{(2\pi)^3\, 2 \bar E_D} \times (2\pi)^6 \left| \psi_D \right|^2 
\left| \mathcal{M}_D \right|^2
\prod_{j=1}^{n_D} \frac{\dd^3 k'_j}{(2\pi)^3 2 E'_{kj}}
\label{Rc} \\ && \times
(2\pi)^3 
\label{Rd} \tau \Delta t\, \bar\delta(\bar E_S - \bar Q - E'_S).
\end{eqnarray}
\end{subequations}
The symbol $\mathcal{M}_S$ denotes the matrix element of the 
source process with $n_S$ particles in the final state, including the 
neutrino. With analogous meanings, $\mathcal{M}_D$ and $n_D$ refer to 
the detection process. In particular, in 
equation~(\ref{R}) we have taken into account all factors $\pi$. 
Equation~(\ref{Rb}) stems from the neutrino propagator in the 
asymptotic limit $L \to \infty$~\cite{GS} and 
is multiplied by the neutrino oscillation probability that 
originates in the phase $\exp (-i m_j^2 L/(2\bar Q)$ of 
equation~(\ref{exp}). Clearly, in the end we have to arrive at an expression 
of the form
\begin{equation}\label{final}
\mathrm{diff}\,\Gamma \times \frac{1}{L^2} P_{\nu_\alpha\to\nu_\beta} \times 
\dd \sigma_D \quad \mbox{with} \quad
P_{\nu_\alpha\to\nu_\beta} = \left| \sum_j U_{\beta j} U_{\alpha j}^* 
e^{-iL m_j^2/(2\bar Q)} \right|^2,
\end{equation}
$\mathrm{diff}\,\Gamma$ being some differential decay rate and 
$\dd \sigma_D$ the infinitesimal cross section of the detection process.
The following steps have to be performed to achieve this aim.
\begin{description}
\item[Neutrino momentum and propagator:]
In our approximation, neutrinos are massless in the source and detection 
process. Moreover, we have to insert the numerator of the neutrino 
propagator that we have left out so far, 
\textit{cf.}\ appendix~\ref{integrations}, that is
\begin{equation}\label{q}
\slashed{q} = \sum_s u(q,s) \bar u(q,s) 
\quad \mbox{with} \quad 
q = \left( \begin{array}{c} \bar Q \\ \vec q \end{array} \right), \quad 
\vec q = \bar Q \vec \ell. 
\end{equation}
In the sum over $s$ only the negative helicity contributes and provides 
the spinors $\bar u(q,-)$ for the neutrino in the final state of the 
source process and $u(q,-)$ for the neutrino in the intial state of the 
detection process. These spinors are not visible in equation~(\ref{R}) 
because they are part of $\mathcal{M}_S$ and $\mathcal{M}_D$, respectively.
\item[Rates:]
In order to obtain a decay rate in the source process, we have to divide 
equation~(\ref{R}) by $\Delta t$. Similarly, we divide by $\tau$ to arrive 
at an event rate in the detection process.
\item[Wave functions:] 
The disturbing presence of the wave functions $\psi_S$ and $\psi_D$ can be 
remedied by assuming that they are 
strongly peaked around $\vec p_\mathrm{in}$ and 
$\vec k_\mathrm{in} = \vec 0$, respectively. Choosing the source process for 
definiteness, the usual trick~\cite{peskin}  
\begin{eqnarray}
\lefteqn{(2\pi)^3 \int \dd^3 p \int \dd^3 p' \,
\delta(\vec p - \vec q - {\vec p}_S^{\,\prime} )\,
\delta( {\vec p}^{\,\prime} - \vec q - {\vec p}_S^{\,\prime} )\, 
\psi_S(\vec p\,) \psi_S^*({\vec p}^{\,\prime}) \cdots}
\nonumber \\
&=& 
(2\pi)^3 \int \dd^3 p \int \dd^3 p' \,
\delta(\vec p - \vec q - {\vec p}_S^{\,\prime} )\,
\delta( {\vec p}^{\,\prime} - \vec p\, ) \,
\psi_S(\vec p\,) \psi_S^*({\vec p}^{\,\prime}) \cdots
\nonumber \\
&=&
(2\pi)^3 \int \dd^3 p\, |\psi_S(\vec p\,)|^2\, 
\delta(\vec p - \vec q - {\vec p}_S^{\,\prime} ) \cdots 
\nonumber \\
&\simeq& 
(2\pi)^3 \delta(\vec p_\mathrm{in} - \vec q - {\vec p}_S^{\,\prime} ) \cdots
\end{eqnarray}
can be applied. In this way we obtain $\delta$-functions for the spatial 
momenta in the source and detection process. 
For the neutrino momentum $\vec q$ see equation~(\ref{q}) and 
the energy of the source particle is given by 
\begin{equation}
\bar E_S = \sqrt{m_S^2 + \vec p_\mathrm{in}^{\,2}}.
\end{equation}
\item[Neutrino in the final state:]
In the source process we need the factor
\begin{equation}
\frac{\dd^3 q}{(2\pi)^3\, 2 \bar Q\rule{0pt}{10pt}} = 
\frac{{\bar Q}^2 \dd \bar Q \dd \Omega}{(2\pi)^3\, 2 \bar Q\rule{0pt}{10pt}}
\end{equation}
for obtaining a decay rate. Since the direction between source and decay 
process is fixed by $\vec \ell$, the result will not be the total decay rate 
but the differential decay rate with respect to the infinitesimal 
solid angle $\dd \Omega$. Note that $L^2 \dd\Omega$ 
corresponds to an infinitesimal 
area at the detection process. Similarly, as we will see below, 
the decay rate will also be differential with respect 
to $\dd \bar Q$. Thus, effectively a factor $\bar Q/(2\,(2\pi)^3)$ is 
inserted and, in the further discussion, we have to remember to 
compensate for it later. 
\item[Neutrino in the initial state:]
In the detection process we need a factor $1/(2\bar Q)$. So the total 
compensating factor is now $4\, (2\pi)^3$.
\item[Neutrino energy:] The neutrino energy has to satisfy 
equation~(\ref{EDQ}). This is achieved by inserting the integration 
$\int \dd \bar Q\, \delta(\bar Q + m_D - E'_D)$
into equation~(\ref{R}). For the differential decay rate with respect to 
$\dd \bar Q$ we simply drop $\int \dd \bar Q$ and arrive at 
$\dd^2 \Gamma/(\dd\Omega \dd \bar Q)$. The energy $\delta$-function is 
attributed to the detection process.
\item[Factors of \boldmath$2\pi$\unboldmath:]
Finally, we collect all numerical factors. These are the total 
compensating factor, the factor from equation~(\ref{Rb}) and that from 
equation~(\ref{Rd}), leading to
\begin{equation}
4 \,(2\pi)^3 \times \frac{\left( 2\pi^2 \right)^2}{(2\pi)^8}
\times (2\pi)^3 = (2\pi)^2.
\end{equation}
These are just the two factors $2\pi$ that are needed in the context of 
energy conservation:
\begin{equation}
2\pi\,\bar\delta(\bar E_S - \bar Q - E'_S) \times 
2\pi\,\delta(\bar Q + m_D - E'_D).
\end{equation}
We find that our formalism automatically leads to the correct 
number of factors $2\pi$.
\end{description}

In summary, we have obtained the differential decay rate  
\begin{eqnarray}
\mathrm{diff}\,\Gamma \equiv 
\lefteqn{\frac{\dd^2 \Gamma}{\dd \Omega \dd \bar Q\rule{0pt}{10pt}}
= \frac{1}{2 \bar E_S} 
\int \prod_{i=1}^{n_S-1} \frac{\dd^3 p'_i}{(2\pi)^3\, 2 E'_{pi}}}
\nonumber \\ && \times
(2\pi)^4 \delta(\vec p_\mathrm{in} - \vec q - {\vec p}_S^{\,\prime} ) 
\,\bar\delta(\bar E_S - \bar Q - E'_S) \left| \mathcal{M}_S \right|^2 
\times \frac{{\bar Q}^2}{(2\pi)^3\, 2\bar Q\rule{0pt}{10pt}}
\label{diffGamma}
\end{eqnarray}
and the detection cross section
\begin{equation}\label{dsigma}
\dd\sigma_D = 
\frac{1}{2 m_D\,2 \bar Q\rule{0pt}{10pt}} \times 
(2\pi)^4 \delta(\vec k_\mathrm{in} + \vec q - {\vec k}_S^{\,\prime} ) 
\,\delta(\bar Q + m_D - E'_D) 
\left| \mathcal{M}_D \right|^2
\prod_{j=1}^{n_D} \frac{\dd^3 k'_i}{(2\pi)^3\, 2 E'_{kj}}.
\end{equation}
The correct energy dimensions of these expressions can be checked with 
\begin{equation}
\dim \left| \mathcal{M}_S \right|^2 = 6 - 2n_S, \quad
\dim \left| \mathcal{M}_D \right|^2 = 4 - 2n_D.
\end{equation}
The integral symbol in 
equation~(\ref{diffGamma}) indicates that in the source process usually the 
final particles are not measured and, therefore, one has to sum 
over all final momenta except the neutrino momentum.

\section{Conclusions}
\label{concl}
In this paper we have introduced a minimal QFT model of neutrino oscillations 
that attempts to take into account the environment in the neutrino production 
and detection process by assuming that production and detection are 
``interrupted'' by collisions with surrounding atoms and molecules. 
In the case of reactor neutrino production and neutrino detection 
the collisions take place in a thermal environment while in the case of 
accelerator neutrinos the charged-pion decay is interrupted when the 
trajectory of the pion intersects the end of the decay tunnel. 
These ``interruptions'' are parameterized in time by 
Heaviside functions such that the source particle has a time interval of 
$\Delta t$ for uninterrupted decay while the detector particle has a 
time interval $\tau$ for undisturbed measurement.\footnote{In~\cite{schwetz} 
Gaussian wave packets are used instead of Heaviside functions and the 
final states are described by Gaussians instead of plane waves. Otherwise the 
topics and results in the present paper have some overlap 
with~\cite{schwetz}.}  
Our main results are the following:
\begin{itemize}
\item
Oscillations occur in space, \emph{not} in time.
\item
In realistic situations there is no decoherence in $\ma$---\textit{cf.}\ 
section~\ref{time}.\footnote{Decoherence takes place only by 
``classical averaging'' at the probability level in the rate of the 
compound process~\cite{schwetz}. This averaging is caused by measurement 
inaccuracies or lack of knowledge of momenta and energies in the final 
state of the detection process and by inaccuracies in the determination of 
$\vec x_S$ and $\vec x_D$.}
\item
Notwithstanding that we work in a QFT model, the resulting 
correlation between the time of the production of the neutrino source particle 
and neutrino detection completely coincides with that of a classical 
consideration---\textit{cf.}\ section~\ref{time}.
\item
Our formalism correctly reproduces the factorization of the event rate 
of the compound production--detection process into the decay rate of the 
source particle, the neutrino oscillation probability and the detection 
cross section---see equations~(\ref{final}), (\ref{diffGamma}) 
and~(\ref{dsigma}).
\end{itemize}
The present notes are a substantial generalization of the discussion 
in~\cite{grimus2019} where roughly speaking the special 
case $T \ll \tau$, $\Delta t\, \Gamma \gg 1$ has been considered.

The following interesting observations concerning $\Delta t$, the time 
interval in which the source particle decays, and $\tau$, the time interval 
when the measurement takes place, can be made:
\begin{itemize}
\item
As discussed in section~\ref{function}, these time intervals prevent 
decoherence in the amplitude.
\item
Since they are arbitrary or unknown, it is gratifying 
that, in good approximation, in the end they drop out---\textit{cf.}\ 
section~\ref{function}. 
\end{itemize}

\vspace{5mm}

\paragraph{Acknowledgement:} The author thanks Thomas Schwetz for 
clarifying discussions.

\newpage

\appendix

\setcounter{equation}{0}
\renewcommand{\theequation}{A.\arabic{equation}}
\section{Integrations}
\label{integrations}
For reasons discussed in the main body of the paper, we omit the matrix 
elements of the weak currents. We also assume that the decay of the 
neutrino source particle produced or present at time $t_D$ is interrupted 
after a time $\Delta t$ by collisions with particles in the environment. 
Therefore, the time window in which the source particle decays has 
a length of $\Delta t$. The decay width of the source particle is denoted by 
$\Gamma$. The uncertainty in the detection time is modelled 
by the function $\Theta_\tau(t_2 - t_D)$ defined in equation~(\ref{Thetatau}). 
With these assumptions the oscillation amplitude is proportional 
to
\begin{subequations}\label{AA}
\begin{eqnarray}
\ma & \propto &
\sum_j U_{\beta j} U_{\alpha j}^* 
\int_{t_S}^{t_S + \Delta t} \dd t_1 \int \dd^3 x_1 
\int_{-\infty}^\infty \dd t_2 \int \dd^3 x_2 
\label{AAa} \\ && \times
\int \dd^3p\, \psi_S(\vec p\,)\,\exp{\left[ -i\,p \cdot (x_1 - x_S) - 
\frac{\Gamma m_S}{2E_S}(t_1 - t_S) + 
i\,p'_S \cdot x_1 \right]} \hphantom{xxx}
\\ && \times
\int \frac{\dd^4 q}{(2\pi)^4}\, 
\frac{e^{-i q \cdot \left( x_2 - x_1 \right)}}{q^2 - m_j^2 + i \epsilon}
\label{AAc} \\ && \times
\int \dd^3k\, \psi_D(\vec k\,)\,\exp{\left[ -i k \cdot (x_2 - x_D)  
+ i\,k'_D \cdot x_2 \right]}
\,\Theta_\tau(t_2 - t_D).
\end{eqnarray}
\end{subequations}
In equation~(\ref{AAa}), $U$ is the CKM lepton mixing matrix and the 
propagator for a neutrino with mass $m_j$ occurs in 
equation~(\ref{AAc}).\footnote{Since we do not consider the 
weak matrix elements for the time being, we also leave out 
the numerator $\slashed{q} + m_j$ of the neutrino propagator.}
We simplify the amplitude by the variable transformation~\cite{ioannisian} 
\begin{equation}
x'_1 = x_1 - x_S, \quad x'_2 = x_2 - x_D
\end{equation}
and obtain
\begin{subequations}\label{A0}
\begin{eqnarray}
\ma & \propto &
e^{i \left( p'_S \cdot x_S + k'_D \cdot x_D \right)} 
\sum_j U_{\beta j} U_{\alpha j}^* 
\int_0^{\Delta t} \dd t'_1 \int \dd^3 x'_1 
\int_{-\infty}^\infty \dd t'_2 \int \dd^3 x'_2 
\\ && \times
\int \dd^3p\, \psi_S(\vec p\,)\,\exp{\left[ -i\, p \cdot x'_1 - 
\frac{\Gamma m_S}{2E_S} t'_1 + i\, p'_S \cdot x'_1 
\right]}
\\ && \times
\int \frac{\dd^4 q}{(2\pi)^4}\, 
\frac{e^{-i q \cdot \left( x'_2 - x'_1 \right)} \times e^{-i q \cdot (x_D - x_S)}}%
{q^2 - m_j^2 + i \epsilon}
\\ && \times
\int \dd^3k\, \psi_D(\vec k\,)\,\exp{\left[ -i k \cdot x'_2  
+ i\,k'_D \cdot x'_2 \right]}
\,\Theta_\tau(t'_2).
\end{eqnarray}
\end{subequations}
Integration over ${\vec x}^{\,\prime}_1$ and ${\vec x}^{\,\prime}_2$
leads to the product of delta functions
\begin{equation}
\delta\,\Big( \vec p - {\vec p}^{\,\prime}_S - \vec q \,\Big) \times 
\delta\left( \vec k - {\vec k}^{\,\prime}_D + \vec q \right).
\end{equation}
The subsequent integrations over $\vec p$ and $\vec k$ result in the 
substitutions
\begin{equation}\label{subs}
\vec p = {\vec p}^{\,\prime}_S + \vec q
\quad \mbox{and} \quad
\vec k = {\vec k}^{\,\prime}_D - \vec q,
\end{equation}
respectively. Integration over $t'_1$ leads to 
\begin{equation}\label{bw}
\frac{\exp \left[ \left( -i \left(E_S - E'_S - q^0 \right) - 
\frac{\displaystyle \Gamma m_S}{\displaystyle 2E_S} \right) \Delta t \right] 
- 1}{-i \left(E_S - E'_S - q^0 \right) - 
\frac{\displaystyle \Gamma m_S}{\displaystyle 2E_S}}.
\end{equation}
Defining
\begin{equation}
\widetilde \Theta_\tau(E) = \int \dd t\, e^{-iEt} \Theta_\tau(t),
\end{equation}
we obtain
\begin{equation}\label{ttheta}
\widetilde \Theta_\tau(E_D - E'_D + q^0) 
\end{equation}
from the integration over $t_2$. 
The energies $E_S$, $E'_S$, $E_D$, $E'_D$ denote the time components of 
$p$, $p'$, $k$, $k'$, respectively.
The remaining two integrations are discussed in the main 
body of the paper.

\setcounter{equation}{0}
\renewcommand{\theequation}{B.\arabic{equation}}
\section{A useful integral}
\label{integral}
We consider the integral
\begin{equation}\label{Ix}
I = 2i \int \dd x \frac{e^{-icx}}{x - a - ib} 
\frac{\sin \left( \lambda x - d \right)}{\lambda x - d}
\quad \mbox{with} \quad \lambda > 0,\;\; b < 0.
\end{equation}
The parameters $a$, $c$ and $d$ are real as well but can be positive or 
negative. Defining
\begin{equation}
z_1 = a + ib \quad \mbox{and} \quad z_2 = \frac{d}{\lambda},
\end{equation}
we note that the integrand has a pole at $z_1$ in the complex plane 
but is analytic at $z_2$. Because of this, we can shift the integration path 
that goes along the $x$-axis into the complex plane 
in the vicinity of $z_2$ and use the residue theorem for the computation of 
$I$. Denoting the thus obtained path in the complex plane by $C$,
we split $I$ into 
\begin{equation}
I = I_1 - I_2 
\quad \mbox{with} \quad
I_1 = \int_C \dd z \frac{e^{-icz}}{z - a- ib} 
\frac{e^{i(\lambda z - d)}}{\lambda z - d},
\;\;
I_1 = \int_C \dd z \frac{e^{-icz}}{z - a - ib} 
\frac{e^{-i(\lambda z - d)}}{\lambda z - d}.
\end{equation}
Now the application of the residue theorem is straightforward. 
For our purpose it is useful to conceive $I$ as a function of $c$.
The result is 
\begin{subequations}\label{I}
\begin{align}
I(c) &= 0 & &\mbox{for} & c &< -\lambda, 
\label{Ia} \\
I(c) &= -\frac{2\pi i}{\lambda} \frac{e^{-icz_2}}{z_1 - z_2}  
\left( 1 - e^{-i(\lambda + c)(z_1 - z_2)} \right) &
&\mbox{for} & |c| &< \lambda, 
\label{Ib} \\
I(c) &= -\frac{2\pi i}{\lambda} \frac{e^{-icz_2}}{z_1 - z_2}  
\left( e^{i(\lambda - c) (z_1 - z_2)} - e^{-i (\lambda + c) (z_1 - z_2)} \right) &
&\mbox{for} & c &> \lambda.
\label{Ic}
\end{align}
\end{subequations}
We emphasize that $I(c)$ is continuous in $c$.

\end{document}